
*****  ATTACHMENT: int3.tex *****

%



\font \titlefont   = cmr10 scaled \magstep2
\font \authorfont  = cmr10 scaled \magstep1

\font \sectionfont = cmr10 scaled \magstep1


\magnification =\magstephalf


\hsize = 17 true cm
\vsize = 24 true cm
\hfuzz=7pt


\newcount \formulanr
\newcount \theoremnr
\newcount \sectionnr
\newcount \remarknr


\formulanr = 0
\theoremnr = 0
\sectionnr = 0
\remarknr  = 0


\def\title#1{\centerline {\titlefont #1} \bigskip}
\def\authors#1{\centerline {\authorfont #1} \bigskip}

\def\section#1{\bigbreak\noindent\advance\sectionnr by 1
                  {\sectionfont \the\sectionnr. #1} \medskip}


\def\references{\bigbreak\centerline{\sectionfont References}
                \bigskip\nobreak}
\def\remark{\smallbreak \noindent \advance\remarknr by 1
            {\bf Remark \the\remarknr. }}
\def\theorem#1{\smallbreak \noindent \advance\theoremnr by 1
                   {\bf Theorem \the\theoremnr. }
                   {\sl #1} \smallbreak \noindent }
\def\proposition#1{\smallbreak \noindent \advance\theoremnr by 1
                   {\bf Proposition \the\theoremnr. }
                   {\sl #1} \smallbreak \noindent }
\def\corollary#1{\smallbreak \noindent \advance\theoremnr by 1
                     {\bf Corollary \the\theoremnr. }
                     {\sl #1} \smallbreak \noindent }
\def\lemma#1{\smallbreak \noindent \advance\theoremnr by 1
                 {\bf Lemma \the\theoremnr. }
                 {\sl #1} \smallbreak \noindent }
\def\proof{\smallbreak \noindent {\rm Proof.} }
\def\endproof{\hfill\nobreak\vskip.1 true cm\nobreak
               \hfill\halmos
               \vskip.1 true cm\goodbreak\noindent}

\def\formula{\global \advance \formulanr by 1 (\the \formulanr)}
\def\form{\eqno\formula}

\def\halmos{\vrule height7pt width5pt}

\def\R{{\bf R}}

\def\a{\alpha}

\def\l{\lambda}
\def\m{\mu}

\def\mod{\hbox{\rm mod}\;}
\def\moda#1{\hbox{\rm mod}_{#1}\;}
\def\u{{\bf u}}

\hrule width 0 pt
\vskip 3 true cm

\title{Multi-Hamiltonian formulation for a class of}
\title{degenerate completely integrable systems}
\bigskip
\authors{Peter Bueken}

\vskip 5 true cm

\noindent{\bf Abstract. }  Generalizing a construction of P. Vanhaecke,
we introduce a large class of degenerate (i.e., associated to a
degenerate Poisson bracket)
completely integrable systems on (a
dense subset of) the space $\R^{2d+n+1}$, called the generalized master
systems.  It turns out that certain
generalized master systems (with different Poisson
brackets and different Hamiltonians)
determine the same Hamiltonian vector fields (and are therefore
different descriptions of the same Hamiltonian system), and that the
Poisson brackets of these systems are compatible.  Consequently,
our class of generalized master systems actually consists of
a (smaller) class of completely integrable systems, and our construction
yields a multi-Hamiltonian structure for these systems.
As an application, we construct a multi-Hamiltonian structure for
the so-called master systems introduced by D. Mumford.

\vfill

\noindent{\bf Keywords: } Completely integrable Hamiltonian
  systems, bi-Hamiltonian structure, multi-Hamiltonian structure,
  compatible Poisson brackets.

\bigskip

\noindent{\bf AMS Classification: } 58F07.

\vskip 2 true cm

\noindent\vbox{\hbox{Address: Katholieke Universiteit Leuven}
               \hbox{\hphantom{Address: }Department of Mathematics}
               \hbox{\hphantom{Address: }Celestijnenlaan 200 B}
               \hbox{\hphantom{Address: }B-3001 Leuven}
               \hbox{\hphantom{Address: }Belgium}}
\medskip
\noindent E-mail: fgaga04@cc1.kuleuven.ac.be

\eject

\baselineskip 20 true pt


\section{Introduction}


It was discovered in the early seventies that certain
well-known partial differential equations
(e.g., the Korteweg-de Vries equation) have a very natural
interpretation as (infinite-dimensional) completely integrable systems,
and that the results from the classical study of finite-dimensional
integrable systems could be used to clarify and to study
the behaviour of these systems.  The discovery of this relation also
caused a revival of the study of (finite-dimensional) completely
integrable systems, which has been undergoing, during the last decades,
a number of significant developments.  On the one hand,
many new examples of (finite-dimensional) completely integrable systems were
constructed (e.g., reductions of infinite-dimensional systems),
and a lot of different techniques
(e.g., algebraic geometric methods) were developed to
study and solve these systems.  On the other hand,
the study of infinite-dimensional integrable systems revealed the
existence of a number of interesting concepts (e.g., recursion
operators) related to the integrability of a system,
leading to a more ``structural'' study of integrable systems, which aims at
a better understanding of integrability and of the structure
of integrable systems.

One of the concepts that evolved from this second line of research
is that of a bi-Hamiltonian structure for a mechanical system.
Roughly speaking, a dynamical system is said to be {\it bi-Hamiltonian}
if it can be written in Hamiltonian form with respect to two different
Poisson structures $\{\cdot,\cdot\}_1$ and $\{\cdot,\cdot\}_2$, i.e.,
  $$\dot{x} = \{x,H_1\}_1 = \{x,H_2\}_2$$
for some functions $H_1,H_2 \in C^\infty(M)$, under the additional
assumption that the brackets $\{\cdot,\cdot\}_1$ and $\{\cdot,\cdot\}_2$
are {\it compatible}, i.e., their sum is again a Poisson bracket.
The existence of different (compatible) Poisson brackets generating the
dynamics of a system was first observed by F. Magri [Ma] in
the case of the Korteweg-de Vries equation, and it was remarked by the
same author that the existence of a bi-Hamiltonian structure for
a mechanical system is closely related to the integrability of the system.
Indeed, if we suppose the Poisson bracket $\{\cdot,\cdot\}_1$ to be
{\sl non-degenerate}, we can define a {\it recursion operator}
  $$N = Q \cdot P^{-1},$$
$P$ and $Q$ being the Poisson tensors associated to the brackets
$\{\cdot,\cdot\}_1$ and $\{\cdot,\cdot\}_2$ respectively.
This recursion operator $N$ can be used (as is shown, for example,
in [MM]) to construct a family of functions $F_i$, which are
first integrals of the Hamiltonian system.
Moreover, these functions turn out to be in involution with respect to
both Poisson brackets,
 $$ \{F_i,F_j\}_1 = \{F_i,F_j\}_2 = 0$$
for all $i,j$, which, in the case of finite-dimensional systems, suffices
to show the integrability of the Hamiltonian system.

We remark however that the precise relation between
bi-Hamiltonian structures and integrability is not completely
understood at this moment. On the one hand, starting
from an infinite-dimensional bi-Hamiltonian system,
the recursion operator construction does not
always yield a sufficient number of first integrals
to prove complete integrability.  Also, in the case of
degenerate systems, (i.e., systems whose associated Poisson structure is
degenerate,) the recursion operator does not exist, and hence the
construction mentioned above cannot be used to prove the integrability
of the system.  On the other hand, it turns out that not all
(finite-dimensional) completely integrable systems admit
a bi-Hamiltonian structure.  In [B], R. Brouzet constructed
a (four-dimensional) non-degenerate
completely integrable system, (i.e., associated to a
non-degenerate Poisson bracket,)
which does not admit a bi-Hamiltonian structure.  The geometry of
non-degenerate completely integrable systems admitting a bi-Hamiltonian
structure was later investigated in [BMT], and the results obtained in
that paper strongly suggest that bi-Hamiltonian formulations for
non-degenerate integrable systems will be rather exceptional.
(To our knowledge, there is only one example of such a system, namely the
Calogero system (see for example [MM]).)

The problem of finding bi-Hamiltonian structures for {\sl degenerate}
completely integrable systems was investigated by many authors,
and turns out to be completely different from the non-degenerate case.
Indeed, a number of different techniques
(e.g., $R$-matrices, master symmetries, reductions of partial
differential equations) were applied successfully to construct
bi-Hamiltonian structures in this situation,
and several important examples of (degenerate) completely integrable systems
(e.g., the Toda lattice, some generalized H\'enon-Heiles potentials and
the so-called master systems) were shown to admit bi-Hamiltonian
structures.  (We refer to [AR1],[AR2],[BR],[D],[Va1] for examples.)
However, a systematic procedure to find such a structure is, to our
knowledge, not known at this moment.

\medskip


Starting from an arbitrary non-zero polynomial $\varphi(x,y)\in\R[x,y]$,
P. Vanhaecke [Va2] constructed a Poisson bracket $\{\cdot,\cdot\}^\varphi$
on $\R^{2d}$ given, in terms of the
standard coordinates $(u_1,\dots,u_d,v_1,\dots,v_d)$ by
  $$\eqalign{& \{u(\l),u_j\}^\varphi = \{v(\l),v_j\}^\varphi = 0,\cr
             & \{u(\l),v_j\}^\varphi = \{u_j,v(\l)\}^\varphi =
                  \varphi(\l,v(\l))
                  \left[{u(\l)\over \l^{d+1-j}}\right]_+ \mod u(\l),\cr}$$
for all $j=1,\dots,d$, where $u(x)=x^d+u_1x^{d-1}+\dots+u_d$ and
$v(x)=v_1x^{d-1}+\dots+v_d$.  (We have denoted by $[R(\l)]_+$ the
polynomial part of the rational function $R(\l)$, and by $P(\l) \mod
u(\l)$ the remainder of the division of $P(\l)$ by $u(\l)$.)
Taking the product of this Poisson manifold with the trivial Poisson
manifold $(\R^{n+1},\{\cdot,\cdot\}_0)$ (with standard coordinates
$c_0,\dots,c_n$ which, in what follows, will be identified
with the coefficients of the polynomial $c(x) = c_0x^n+\dots+c_n$),
one obtains a Poisson bracket
$\{\cdot,\cdot\}^\varphi_c$ on $\R^{2d+n+1}$ which has the functions
$c_0,\dots,c_n$ as Casimir functions.

Next, we consider deformation families of polynomials of the form
  $$F(x,y,c) = F(x,y) - x^d c(x),\qquad c(x) = c_0x^n+\dots+c_n,$$
where $F(x,y)$ is a polynomial depending explicitly on $y$.
It was shown in [Va2] that the (independent)
coefficients $H_1,\dots,H_d$ of the polynomial
  $$H(\l) = H_1\l^{d-1}+\dots+H_d = (F(\l,v(\l))-\l^dc(\l))\mod u(\l)$$
are in involution with respect to the Poisson bracket
$\{\cdot,\cdot\}^\varphi_c$ (for every polynomial $\varphi$),
and hence determine (together with the
Casimir functions $c_0,\dots,c_n$,) a completely integrable
system on the Poisson manifold $(\R^{2d+n+1},\{\cdot,\cdot\}^\varphi_c)$.

The first aim of this paper is to show that the functions in the set
  $$I=\{c_0,\dots,c_n,H_1,\dots,H_d\}$$
actually Poisson-commute with respect to a much larger class of
Poisson structures $\{\cdot,\cdot\}^\varphi_\a$ (indexed by the polynomials
$\varphi$ and the integers $\a\le n+1$).  Intuitively speaking,
these new brackets are constructed from the Poisson
bracket $\{\cdot,\cdot\}^\varphi_c$ by ``choosing another
set of Casimir functions''.  More precisely, the construction works as
follows.  For any integer $\a\le n+1$, the polynomial
  $$G(\l) = \l^d c(\l) + (F(\l,v(\l)) - \l^d c(\l)) \mod u(\l)$$
(which has the elements of $I$ as its coefficients)
can be written in the form
  $$G(\l) = f_0 \l^{d+n} + \dots + f_{n-\a}\l^{d+\a}
            + \l^\a(g_1\l^{d-1}+\dots +g_d)
            + f_{n-\a+1} \l^{\a-1}+\dots +f_n,$$
(i.e., we make a special partition
  $$P=\{\{f_0,\dots,f_n\},\{g_1,\dots,g_d\}\}$$
of the set $I$).  The mapping
  $$\phi : \R^{2d+n+1}\to \R^{2d+n+1}: (u,v,c)\to (u,v,f(u,v,c))$$
associated to the functions $f_0,\dots,f_n$ is a bijection on (a dense
subset $M$ of) $\R^{2d+n+1}$.  Pushing forward the Poisson bracket
$\{\cdot,\cdot\}^\varphi_c$ with the mapping $\phi^{-1}$, we obtain a
new Poisson bracket $\{\cdot,\cdot\}^\varphi_\a$ (on $M$)
which has the functions $f_0,\dots,f_n$ as Casimir functions.
The remaining (independent) functions $g_1,\dots,g_d$
are in involution with respect to
the new Poisson bracket $\{\cdot,\cdot\}^\varphi_\a$,
and determine a $d$-dimensional
completely integrable system $(M,\{\cdot,\cdot\}^\varphi_\a,g)$.

In the special case where $\varphi=1$, $\a=0$ and
  $$F(x,y) = y^2+f(x),$$
these completely integrable systems turn out to be the master systems
introduced by D. Mumford [M] and P. Vanhaecke [Va1], and all systems
$(M,\{\cdot,\cdot\}^\varphi_\a,g)$ constructed above
will therefore be called {\it generalized master systems}.

It is clear from the construction given above that (for a fixed
polynomial $F(x,y)$) all generalized master systems are different in the
sense that they are defined by different Poisson brackets and different
Hamiltonians.  However, it turns out that not all these systems are
different in the sense that they have different Hamiltonian vector
fields.  To show this, we will compute the Hamiltonian vector fields of
the generalized master systems, and we will search for different systems
(i.e., different Poisson structures and Hamiltonians) yielding the same
Hamiltonian vector fields.  More precisely, we will consider the
generalized master system $(M,\{\cdot,\cdot\}^\varphi_\a,g)$, and we
will search for other partitions $P'$ (associated to the integer
$\a'=\a+i\le n+1$) and other polynomials $\varphi'$ such that the
Hamiltonian vector fields of the generalized master system
$(M,\{\cdot,\cdot\}^{\varphi'}_{\a'},g')$ coincide with those of the
first system, i.e., such that
  $$\{\cdot,g_a\}^\varphi_\a = \{\cdot,g_a'\}^{\varphi'}_{\a+i},$$
for all $a=1,\dots,d$.  We will show that this condition is satisfied
when $\varphi' = x^i \varphi$, thus proving that
all Hamiltonian vector fields of the generalized master system
$(M,\{\cdot,\cdot\}^\varphi_\a,g)$ can be written in Hamiltonian form with
respect to the Poisson brackets
$\{\cdot,\cdot\}^{x^i\varphi}_{\a+i}$,
for all $i$ such that $\a+i\le n+1$.
We will also show that all these Poisson brackets are compatible,
leading to a
multi-Hamiltonian formulation for ``all'' generalized master systems.
(Strictly speaking, the above argument shows that our large class of
generalized master systems consists of a (smaller) class of completely
integrable systems (which turns out to be exactly the class introduced
in [Va2], if we allow $\varphi$ to be a special rational function),
together with the systems determined by their
multi-Hamiltonian formulation, and from this (stricter) point of view,
our construction yields a
multi-Hamiltonian structure for the smaller class of systems introduced
in [Va2].)

Finally, we remark that the construction of the
Poisson bracket $\{\cdot,\cdot\}^\varphi_\a$
can easily be adapted to the case where $P$ is an {\sl arbitrary}
partition
  $$P=\{\{f_0,\dots,f_n\},\{g_1,\dots,g_d\}\}$$
of the set $I$, yielding a new Poisson bracket
$\{\cdot,\cdot\}^\varphi_f$ which has the functions
$f_0,\dots,f_n$ as Casimir functions.  The remaining functions
$g_1,\dots,g_d$ again Poisson-commute with respect to this Poisson bracket
and hence determine a completely integrable system,
also called a generalized master system.  The procedure given above can
be generalized to this larger class of completely integrable systems,
again showing that our collection of systems consists of a (smaller) class
of integrable systems, together with their multi-Hamiltonian
formulation.  (We remark however, that this class of (different)
systems is larger than the one considered in [Va2], i.e., we have actually
constructed {\sl new} systems, in the sense that they have different
Hamiltonian vector fields.)

We have already remarked that, putting $\varphi=1$, $\a=0$ and
$F(x,y)=y^2+f(x)$ in our description, we obtain the well-known (odd and
even) master systems.  In the two-dimensional case, the existence of a
multi-Hamiltonian structure for these master systems was shown by P.
Vanhaecke in [Va1].  However, the construction given in [Va1] involves a
heuristic choice of the Casimir functions, and uses a trial-and-error
method to compute the components of the different Poisson brackets
involved in this multi-Hamiltonian structure.  As an application of our
results, we will give a complete description of
the multi-Hamiltonian structure of the two-dimensional odd master system.
We will compute the Hamiltonian vector fields of this system, and we will
construct the different (compatible) Poisson brackets involved in its
multi-Hamiltonian structure.

The paper is organized as follows.  In Section 2, we collect some
preliminary material concerning bi-Hamiltonian structures and integrable
systems, and we fix some notation which will be used
in the rest of the paper.
In Section 3 we start by briefly reviewing the results from [Va2].
Further, we will also introduce the generalized master systems, and we
will prove the integrability of these systems.
In Section 4, we describe the multi-Hamiltonian structure
for these generalized master systems.
Finally, in Section 5 we use the theory developed in this paper to
construct a multi-Hamiltonian structure for
the odd master system.


\section{Preliminaries}


A {\it Poisson manifold} $(M,\{\cdot,\cdot\})$ is a smooth manifold $M$,
endowed with an anti-symmetric $\R$-bilinear mapping
  $$\{\cdot,\cdot\} : C^\infty(M) \times C^\infty(M) \to C^\infty(M)$$
called a {\it Poisson bracket}, which is a derivation in
both of its arguments, and which satisfies the {\it Jacobi identity}
  $$ \{\{f,g\},h\} + \{\{g,h\},f\} + \{\{h,f\},g\} = 0$$
for all $f,g,h \in C^\infty(M)$.  Two functions $f,g\in C^\infty(M)$ are
said to {\it Poisson-commute}, or to be {\it in involution},
if their Poisson bracket vanishes, i.e.,
  $$\{f,g\} = 0.$$
A function $f$ which Poisson-commutes with {\sl all} functions $g\in
C^\infty(M)$ is called a {\it Casimir function}.  As
the Poisson bracket $\{\cdot,\cdot\}$ is a
derivation in its arguments, every function $f \in C^\infty(M)$
determines a vector field $X_f$ on $M$, called the {\it Hamiltonian vector
field} generated by $f$, by the relation
  $$X_f(g) = \{g,f\}$$
for all $g \in C^\infty(M)$.  If $f$ generates a non-zero vector field,
(i.e., $f$ is not a Casimir function,) $f$ is said to be a {\it
Hamiltonian},
and the triple $(M,\{\cdot,\cdot\},f)$ is called a {\it Hamiltonian system}.

Let $(M,\{\cdot,\cdot\})$ be a Poisson manifold of dimension $n$ which
admits $k$ (independent) Casimir functions (i.e., it is of rank
$2d=n-k$).  A Hamiltonian system $(M,\{\cdot,\cdot\},H)$ on this Poisson
manifold is said to be {\it Liouville integrable}
or {\it completely integrable} if
(apart from the Casimir functions $C_1,\dots,C_k$) it admits $d$
independent invariant functions
(sometimes called the {\it Hamiltonians} of the system)
$H_1=H,H_2,\dots,H_d$, which are in
involution with respect to the Poisson bracket $\{\cdot,\cdot\}$.
We will frequently denote such an integrable system by
$(M,\{\cdot,\cdot\},h)$, where $h=(H_1,\dots,H_d)$ is a set of
Hamiltonians of the system.

Two Poisson brackets $\{\cdot,\cdot\}_1$ and $\{\cdot,\cdot\}_2$ on a
manifold $M$ are said to be {\it compatible} if their sum or, equivalently,
any linear combination
  $$\{\cdot,\cdot\}_1 + \l \{\cdot,\cdot\}_2,$$
is again a Poisson bracket.
More generally, a collection of Poisson
brackets are said to be compatible if any linear combination of them is
again a Poisson bracket or, equivalently, if they are pairwise
compatible.

A Hamiltonian system $(M,\{\cdot,\cdot\}_1,H_1)$ is said to be
{\it multi-Hamiltonian} if its Hamiltonian vector field
  $$X_{H_1} = \{\cdot,H_1\}_1$$
can be written in Hamiltonian form with respect to
a set of {\sl compatible} Poisson brackets, i.e.,
  $$\{\cdot,H_1\}_1 = \{\cdot,H_2\}_2 = \dots = \{\cdot,H_n\}_n$$
for some functions $H_2,\dots,H_n \in C^\infty(M)$.
In the special case where $n=2$, we
say that the system is {\it bi-Hamiltonian}.

In the rest of this paper, we will use the following notational
conventions.  Let $u(x)$ be a (generic) polynomial of degree $d$.
Then (for a fixed integer $\a$) any polynomial $P(x)$ can be written,
in a unique way, as
  $$P(x) = Q(x) u(x) + x^\a R(x),$$
where $\deg R(x) < d$.  We will denote
  $$Q(x) = \left[ {P(x) \over u(x)} \right]_\a.$$
Further, we define
  $$ x^\a R(x) = P(x) - u(x) \left[ {P(x)\over u(x)} \right]_\a
      = P(x) \moda{\a} u(x),$$
and we say that
  $$ {P(x)\over u(x)} - \left[ {P(x)\over u(x)} \right]_\a
      = \left[ {P(x) \over u(x)} \right]_\a^-.$$
In the case where $\a=0$ we will, in analogy with the notation in [Va2],
denote
  $$ P(x) \moda{0} u(x) = P(x) \mod u(x),$$
and
  $$ \left[ {P(x)\over u(x)} \right]_0=\left[ {P(x)\over u(x)}
  \right]_+,
  \quad
   \left[ {P(x)\over u(x)} \right]_0^-=\left[ {P(x)\over u(x)}
   \right]_-.$$


\section{Construction of generalized master systems}


In [Va2], P. Vanhaecke constructs, starting from any polynomial
$F(x,y) \in \R[x,y]\setminus \R[x]$, a set of $d$ independent functions
$H_1,\dots,H_d$.  These functions turn out to be in involution with
respect to a large class of Poisson brackets $\{\cdot,\cdot\}^\varphi$
on the space $\R^{2d}$, and hence they determine a family of completely
integrable Hamiltonian systems on this space.
More generally, considering a deformation family
  $$F(x,y,c), \quad c=(c_1,\dots,c_k)\in\R^k,$$
of polynomials, one can construct a new set
  $$I=\{c_1,\dots,c_k,H_1,\dots,H_d\}$$
of independent functions, which again turn out to Poisson-commute with
respect to a family of Poisson brackets $\{\cdot,\cdot\}^\varphi_c$
on $\R^{2d+k}$
(which have the functions $c_1,\dots,c_k$ as Casimir functions)
and therefore determine a collection of completely integrable
systems on the space $\R^{2d+k}$.  The aim of this section is to show
that the functions in the set $I$ actually Poisson-commute with respect
to a much larger class of Poisson brackets on (a dense subset $M$ of) the
space $\R^{2d+k}$.  Indeed, it will turn out that, choosing a
``suitable''
set $f_1,\dots,f_k$ of functions on the space $\R^{2d+k}$, we can
construct a Poisson bracket $\{\cdot,\cdot\}^\varphi_f$
on a dense subset $M \subset \R^{2d+k}$, which has the functions
$f_1,\dots,f_k$ as Casimir functions.  The functions in the set $I$ are
in involution with respect to this new bracket, and hence
determine a (new) completely integrable system on $M$.

We start this section by briefly reviewing the construction given in
[Va2].  To this purpose,
we consider the space $\R^{2d}$ as the space of pairs of polynomials
  $$\eqalign{ u(\l) & = \cr v(\l) & = \cr}
    \eqalign{ \l^d+ & u_1\l^{d-1}+\dots+u_d,\cr
              {}    & v_1\l^{d-1}+\dots+v_d,\cr}$$
i.e., the coefficients of the polynomials $u,v$ serve as
coordinates on $\R^{2d}$.  The following result
then gives explicit
expressions (in terms of the coordinates $u_i$, $v_i$ on $\R^{2d}$)
for a family of Poisson brackets, indexed by the space of
(non-zero) polynomials in two variables:
  \proposition{Let $\varphi(x,y)$ be a non-zero polynomial in $\R[x,y]$.
    Then the bracket $\{\cdot,\cdot\}^\varphi$,
    given in terms of the coordinates $u_i$, $v_i$ by
      $$\eqalign{& \{u(\l),u_j\}^\varphi = \{v(\l),v_j\}^\varphi = 0,\cr
                 & \{u(\l),v_j\}^\varphi = \{u_j,v(\l)\}^\varphi =
                      \varphi(\l,v(\l))
                      \left[{u(\l)\over \l^{d+1-j}}\right]_+ \mod u(\l),\cr}
         \form$$
    $j=1,\dots,d$, defines a (polynomial) Poisson structure
    on the space $\R^{2d}$.  Moreover, all these Poisson brackets
    $\{\cdot,\cdot\}^\varphi$ are compatible.}
\noindent The next step is the construction, for every Poisson bracket
given by (1), of a collection of completely integrable systems on
$\R^{2d}$.  This is the purpose of the following
  \proposition{Let $F(x,y)\in \R[x,y]\setminus\R[x]$.  Then the
    coefficients $H_1,\dots,H_d$ of the polynomial
      $$H(\l) = H_1\l^{d-1}+\dots+H_d = F(\l,v(\l)) \mod u(\l) \form $$
    define, for any non-zero polynomial $\varphi(x,y)$, a completely
    integrable system on the Poisson manifold
    $(\R^{2d},\{\cdot,\cdot\}^\varphi)$ with polynomial invariants.}

The previous construction can be extended
to obtain completely integrable systems
on certain product manifolds $\R^{2d}\times D$.
We will now describe this procedure.
However, we will restrict our attention to the case where $D=\R^k$,
as the more general situation will not be considered
in the rest of this paper.

We start by defining a (trivial) Poisson bracket $\{\cdot,\cdot\}_0$ on
$D=\R^k$ by putting
  $$\{f,g\}_0 = 0$$
for all $f,g \in C^\infty(\R^k)$.
Letting $(c_1,\dots,c_k)$ denote the standard coordinates on $\R^k$ and
considering the space $\R^{2d+k}$ as the product of the Poisson
manifolds $(\R^{2d},\{\cdot,\cdot\}^\varphi)$ and
$(\R^k,\{\cdot,\cdot\}_0)$, we obtain
a (product) Poisson structure $\{\cdot,\cdot\}^\varphi_c$
on $\R^{2d+k}$, which has the
coordinate functions $c_1,\dots,c_k$ as Casimir functions.
Suppose that we are given a deformation family
of polynomials, i.e., a family of polynomials
  $$F(x,y,c_1,\dots,c_k)$$
depending on $k$ parameters $c=(c_1,\dots,c_k) \in \R^k$.
(In the more general case, one takes parameters in the manifold $D$.)
Then we have the following
\proposition{The coefficients $H_1,\dots,H_d$ of the polynomial
    $$H(\l) = F(\l,v(\l),c_1,\dots,c_k) \mod u(\l)$$
  are in involution with respect to the Poisson bracket
  $\{\cdot,\cdot\}^\varphi_c$, and hence
  determine a completely integrable Hamiltonian system
  $(\R^{2d+k},\{\cdot,\cdot\}^\varphi_c,h)$,
  $h=(H_1,\dots,H_d)$ on the (degenerate) Poisson manifold
  $(\R^{2d+k},\{\cdot,\cdot\}^\varphi_c)$.}


Let us suppose that $\varphi(x,y)$ is a polynomial and that
$F(x,y,c)$, $c=(c_1,\dots,c_k)\in \R^k$ is a deformation
family of polynomials, and let
$(\R^{2d+k},\{\cdot,\cdot\}^\varphi_c,h)$ be the completely
integrable Hamiltonian system given by Proposition 3.
We will now show how one can construct, starting from the
system $(\R^{2d+k},\{\cdot,\cdot\}^\varphi_c,h)$,
a large collection of other completely integrable
systems on the space $\R^{2d+k}$ (or, more precisely, on a dense subset
$M$ of $\R^{2d+k}$,) which have the same set of invariants.
Roughly speaking, this construction works as follows:
choosing a ``suitable'' set of functions $f_1,\dots,f_k$ in
$C^\infty(\R^{2d+k})$, we construct a new Poisson bracket
$\{\cdot,\cdot\}^\varphi_f$ (on $M$) which has $f_1,\dots,f_k$ as Casimir
functions.  We will show that the invariants
$c_1,\dots,c_k,H_1,\dots,H_d$ of the original system
$(\R^{2d+k},\{\cdot,\cdot\}^\varphi_c,h)$ Poisson-commute with
respect to $\{\cdot,\cdot\}^\varphi_f$ and therefore determine a
completely integrable system.

We start by constructing the new Poisson bracket
$\{\cdot,\cdot\}^\varphi_f$.  To this purpose,
we consider $k$ functions
  $$f_1(c,h(u,v,c)),\dots,f_k(c,h(u,v,c))$$
such that the mapping
  $$\phi : \R^{2d+k}\to \R^{2d+k}: (u,v,c)\mapsto (u,v,f(u,v,c))$$
(where we denoted $f=(f_1,\dots,f_k)$)
is a bijection on a dense subset $M \subset \R^{2d+k}$.
Pushing forward the (restriction to $M$ of the)
Poisson structure $\{\cdot,\cdot\}^\varphi_c$ with
the mapping $\phi^{-1}$, we
obtain a new Poisson bracket $\{\cdot,\cdot\}^\varphi_f$ on $M$.
By construction, the mapping $\phi$ is a Poisson morphism
and hence the functions $f_i(c,h(u,v,c))$, $i=1,\dots,k$ are
Casimir functions for this new Poisson bracket.  The following theorem
states that the (restriction to $M$ of the)
invariants $c_1,\dots,c_k,H_1,\dots,H_d$ of the system
$(\R^{2d+k},\{\cdot,\cdot\}^\varphi_c,h)$ Poisson-commute with
respect to this new bracket $\{\cdot,\cdot\}^\varphi_f$.

\theorem{Let $f_1(c,h(u,v,c)),\dots,f_k(c,h(u,v,c))$ be $k$ functions
  such that the mapping
    $$\phi : \R^{2d+k}\to\R^{2d+k}:(u,v,c) \mapsto (u,v,f(u,v,c))$$
  is bijective on a dense subset $M \subset \R^{2d+k}$, and let
  $\{\cdot,\cdot\}^\varphi_f$ be the Poisson bracket associated to
  $f=(f_1,\dots,f_k)$ as above.  Then the
  invariant functions $c_1,\dots,c_k,H_1,\dots,H_d$
  of the system $(\R^{2d+k},\{\cdot,\cdot\}^\varphi_c,h)$
  are in involution with respect to the Poisson
  bracket $\{\cdot,\cdot\}^\varphi_f$ and hence determine a completely
  integrable Hamiltonian system on the Poisson manifold
  $(M,\{\cdot,\cdot\}^\varphi_f)$.}
\proof
  For the sake of simplicity, we will denote
  $\u=(u_1,\dots,u_d,v_1,\dots,v_d)$, and
  we write the Jacobian
  matrix $J(f,c)$ of the functions $f_i$ with respect to the variables
  $c_j$ by $f_c$.

  As the functions $c_1,\dots,c_k$ are Casimir functions of the Poisson
  bracket $\{\cdot,\cdot\}^\varphi_c$, the Poisson
  matrix (with respect to the coordinates $(\u,c)$)
  of the bracket $\{\cdot,\cdot\}^\varphi_c$ is of the form
    $$\pmatrix{A&0\cr0&0},$$
  where $A$ is the (skew-symmetric) $(2d\times 2d)$-matrix
  given by $\{\u,\u\}^\varphi$.

  Next, we compute the Poisson matrix of the bracket
  $\{\cdot,\cdot\}^\varphi_f$ with respect to the coordinates $(\u,c)$.
  As $\phi$ is a Poisson morphism,
    $$\{\u,\u\}^\varphi_f = \{\u,\u\}^\varphi_c = \{\u,\u\}^\varphi,$$
  and hence its Poisson matrix is of the form
    $$\pmatrix{ A&B\cr -{}^tB&C\cr}.$$
  Taking into account that
  the functions $f_i(c,h(\u,c))$, $i=1,\dots,k$ are Casimir
  functions of the bracket $\{\cdot,\cdot\}^\varphi_f$,
  we obtain that
    $$\pmatrix{A&B\cr -{}^tB&C\cr} \pmatrix{f_\u \cr f_c\cr} = 0.\form $$
  By assumption, $\det f_c \neq 0$ on $M$ and (3) yields
    $$B=-Af_\u f_c^{-1},\qquad C={}^tf_c^{-1} {}^tf_\u A f_\u f_c^{-1}.
      \form$$
  As the functions $f_i$ only depend on $c_j$ and $H_j$ (which
  Poisson-commute with respect to the bracket
  $\{\cdot,\cdot\}^\varphi_c$), we see that
    $$\{f,f\}^\varphi_c = {}^tf_\u A f_\u = 0,$$
  and hence
    $$C = 0.\form$$
  Summarizing, the Poisson matrix of the bracket
  $\{\cdot,\cdot\}^\varphi_f$
  is of the form
    $$\pmatrix{A & -Af_\u f_c^{-1}\cr
               -{}^tf_c^{-1} {}^tf_\u A & 0\cr}.\form$$
  Using this expression, we compute that
    $$\eqalign{ \{h,h\}^\varphi_f = & \pmatrix{ {}^th_\u & {}^th_c\cr}
                                 \pmatrix{A & -Af_\u f_c^{-1}\cr
                                         -{}^tf_c^{-1} {}^tf_\u A & 0\cr}
                                  \pmatrix{h_\u \cr h_c \cr} \cr
                             = & \{h,h\}^\varphi_c
                                 - {}^th_c {}^t f_c^{-1}
                                     \{f,h\}^\varphi_c
                                 - \{h,f\}^\varphi_c f_c^{-1} h_c,\cr}$$
  and as all the functions $f_i$ and $H_j$ are in involution with
  respect to $\{\cdot,\cdot\}^\varphi_c$, this proves that the $H_i$
  Poisson-commute with respect to $\{\cdot,\cdot\}^\varphi_f$.  Analogous
  computations yield that $\{c,h\}^\varphi_f=0$ and
  $\{c,c\}^\varphi_f=0$.  Hence, we obtain (apart from the Casimir
  functions $f_1,\dots,f_k$) a set of $d$ additional independent
  functions in involution, which concludes the proof of the theorem.
\endproof

In the rest of this section, we discuss an application of Theorem 4
which will play an important role in what follows.  We will consider
special deformation families of polynomials.  Inspired by our study of
the master systems (described in Section 5), we make a very simple
and natural choice for the functions $f=(f_1,\dots,f_k)$.  It turns out
that the mapping $\phi$ associated to $f$ is bijective on a
dense subset $M\subset\R^{2d+k}$, and applying Theorem 4 we obtain a
completely integrable system, called a {\it generalized master system},
on the Poisson manifold $(M,\{\cdot,\cdot\}^\varphi_f)$.

We start by fixing an integer $d\ge 1$, a (non-zero)
polynomial $\varphi(x,y)$ and a polynomial
  $$F(x,y)\in \R[x,y]\setminus\R[x],$$
and we consider the deformation family
  $$F(x,y,c_0,\dots,c_n) = F(x,y) - x^d c(x), \quad (c_0,\dots,c_n)\in
  \R^{n+1}, \form$$
where $c(x) = c_0 x^n + \dots + c_n$.
Further, we denote by $(\R^{2d+n+1},\{\cdot,\cdot\}^\varphi_c,h)$
the completely integrable system
constructed in Proposition 3, and by $I$ the set
  $$I=\{c_0,\dots,c_n,H_1,\dots,H_d\}$$
of invariants of this system.  Finally, we also introduce the polynomial
  $$G(\l) = \l^d c(\l) + H(\l) = \l^d c(\l) + (F(\l,v(\l)) - \l^d c(\l))\mod
    u(\l),\form$$
whose coefficients are given by the elements of the set $I$.

Choosing any integer $\a \le n+1$, we can write the polynomial $G(\l)$
in the form
  $$G(\l) = f_0 \l^{d+n} + \dots + f_{n-\a}\l^{d+\a}
            + \l^\a(g_1\l^{d-1}+\dots +g_d)
            + f_{n-\a+1} \l^{\a-1}+\dots +f_n,\form$$
and we obtain a partition
  $$P=\{\{f_0,\dots,f_n\},\{g_1,\dots,g_d\}\}$$
of the set $I = \{c_0,\dots,c_n,H_1,\dots,H_d\}$.
Our aim is to apply Theorem 4, where $f=(f_0,\dots,f_n)$ is given by
the partition $P$ (or, equivalently, by the integer $\a$).
It is clear that, in order to do this, we have to show that the
mapping
  $$\phi : \R^{2d+n+1}\to \R^{2d+n+1} : (u,v,c)\mapsto (u,v,f(u,v,c))$$
associated to $f=(f_0,\dots,f_n)$
is bijective on a dense subset $M\subset \R^{2d+n+1}$.

It is easy to see that (8) implies
  $$G(\l) = F(\l,v(\l))\mod u(\l) + u(\l) \left[ {\l^d c(\l)\over
  u(\l)}\right]_+.\form$$
On the other hand, denoting
  $$\eqalign{f(\l) = & f_0 \l^{d+n} + \dots + f_{n-\a}\l^{d+\a}
                     + f_{n-\a+1} \l^{\a-1}+\dots f_n,\cr
             g(\l) = & g_1\l^{d-1}+\dots+g_d,\cr}$$
(9) takes the form
  $$G(\l) = f(\l) + \l^\a g(\l).\form$$
Comparing (10) and (11) we obtain
  $$\left[ {\l^d c(\l) \over u(\l)} \right]_+
    =\left[ {f(\l) - (F(\l,v(\l))\mod u(\l)) \over u(\l)} \right]_\a,$$
which yields
  $$c(\l) = \left[ {u(\l)\over \l^d}
                   \left[ {f(\l) - (F(\l,v(\l))\mod u(\l)) \over u(\l)}
                          \right]_\a
                 \right]_+.\form$$
The right hand side of this equation is defined for a generic polynomial
$u(\l)$.  More precisely, it is defined for all polynomials $u(\l)$
in the case where $\a=0$, and on the subspace $u_d\neq 0$ in the other cases.
Hence, the mapping $\phi$ is bijective on the dense
subset $M = \R^{2d+n+1}(\setminus \{u_d=0\})$.  As an immediate consequence of
Theorem 4, we obtain the following
\corollary{Let $d\ge 1$ be an integer and $\varphi(x,y)$ any non-zero
  polynomial in $\R[x,y]$.  Further, let
    $$F(x,y,c) = F(x,y) - x^d c(x), \quad c(x) = c_0x^n + \dots + c_n,$$
  be a deformation family of polynomials and denote by
    $$G(\l) = \l^d c(\l) + (F(\l,v(\l))-\l^d c(\l))\mod u(\l)$$
  the polynomial whose coefficients are given by the set
    $$I=\{c_0,\dots,c_n,H_1,\dots,H_d\}$$
  of invariants of the system $(\R^{2d+n+1},\{\cdot,\cdot\}^\varphi_c,h)$.
  For every $\a\le n+1$, the partition
    $$P=\{\{f_0,\dots,f_n\},\{g_1,\dots,g_d\}\}$$
  of the set $I$, given by
    $$G(\l) = f_0 \l^{d+n} + \dots + f_{n-\a}\l^{d+\a}
              + \l^\a(g_1\l^{d-1}+\dots +g_d)
              + f_{n-\a+1} \l^{\a-1}+\dots +f_n,$$
  determines a completely integrable Hamiltonian system
  $(M,\{\cdot,\cdot\}^\varphi_\a,g)$ on the dense subset
  $M = \R^{2d+n+1}(\setminus \{u_d=0\})$,
  called a {\it generalized master system}. }


\section{Multi-Hamiltonian formulation for the generalized master systems}


The main purpose of this section is to construct a multi-Hamiltonian
formulation for the generalized master systems
$(M,\{\cdot,\cdot\}^\varphi_\a,g)$ introduced in Corollary 5.
More precisely, we will prove (the slightly stronger result)
that {\sl all} Hamiltonian vector fields
$X_{g_a}$, $a=1,\dots,d$, (not only $X_{g_1}$)
of the generalized master system
$(M,\{\cdot,\cdot\}^\varphi_\a,g)$ can be written in
Hamiltonian form with respect to a set of compatible Poisson brackets.
The first step in our construction
is to determine explicit expressions for the
Hamiltonian vector fields $X_{g_a} = \{\cdot,g_a\}^\varphi_\a$,
$a=1,\dots,d$, of the generalized master system with respect
to the coordinates $(u,v,c)$.
Next, we show that the Hamiltonian vector fields
  $$X_{g_a} = \{\cdot,g_a\}^\varphi_\a, \quad a=1,\dots,d,$$
of this system coincide with the Hamiltonian vector fields
of a number of other (generalized master) systems
$(M,\{\cdot,\cdot\}^{\varphi'}_{\a'},g')$
(for particular choices of the polynomial
$\varphi'$ and of the integer $\a'$), i.e.,
  $$\{\cdot,g_a\}^\varphi_\a = \{\cdot,g_a'\}^{\varphi'}_{\a'},$$
for all $a=1,\dots,d$.  As an immediate consequence of this fact,
the Hamiltonian vector fields of the generalized master system
$(M,\{\cdot,\cdot\}^\varphi_\a,g)$ can be written
in Hamiltonian form with
respect to a set of Poisson brackets $\{\cdot,\cdot\}^{\varphi'}_{\a'}$
(for certain $\varphi'$ and $\a'$).  Finally, we will show that
any pair of these Poisson brackets are compatible, thus concluding our
construction of a multi-Hamiltonian formulation for the generalized master
systems.

We start our construction by computing
explicit expressions for the Hamiltonian
vector fields of the generalized master system
$(M,\{\cdot,\cdot\}^\varphi_\a,g)$ with respect to the
coordinates $(u,v,c)$.  This is the purpose of the following
\theorem{Let $F(x,y,c)$ be a deformation family of polynomials as above.
  Further, suppose that $\a\le n+1$ is an integer and let
    $$P=\{\{f_0,\dots,f_n\},\{g_1,\dots,g_d\}\}$$
  be the partition of the set
    $$I=\{c_0,\dots,c_n,H_1,\dots,H_d\}$$
  associated to $\a$, i.e., the partition given by
    $$G(\l) = f_0 \l^{d+n} + \dots + f_{n-\a}\l^{d+\a}
              + \l^\a(g_1\l^{d-1}+\dots +g_d)
              + f_{n-\a+1} \l^{\a-1}+\dots +f_n.\form$$
  The Hamiltonian vector fields $X_{g_a}=\{\cdot,g_a\}^\varphi_\a$
  of the generalized master system $(M,\{\cdot,\cdot\}^\varphi_\a,g)$
  are completely determined by (the coefficients in $\m$ of)
    $$\eqalign{
      X_{G(\m)} u(\l) & = {\partial F\over\partial y}(\m,v(\m))
                          \{u(\l),v(\m)\}^\varphi \moda{\a} u(\m),\cr
      X_{G(\m)} v(\l) & = \left[{F(\m,v(\m))-\m^d c(\m)\over u(\m)}
                          \right]_+ \{u(\l),v(\m)\}^\varphi \moda{\a}
                          u(\m).\cr }\form$$}
\proof
  As the functions $f_0,\dots,f_n$ are Casimir functions
  of the Poisson bracket $\{\cdot,\cdot\}^\varphi_\a$, (13) yields
    $$\{\cdot,G(\m)\}^\varphi_\a
      = \m^\a ( \{\cdot,g_1\}^\varphi_\a \m^{d-1} + \dots +
                \{\cdot,g_d\}^\varphi_a).\form$$
  On the other hand, we can rewrite $G(\m)$ as
    $$G(\m) = F(\m,v(\m))
              - u(\m) \left[ {F(\m,v(\m)) - \m^d c(\m)\over u(\m)}
                      \right]_+,$$
  which implies that
    $$\eqalign{ \{\cdot,G(\m)\}^\varphi_\a
        = & \{\cdot,v(\m)\}^\varphi_\a{\partial F\over\partial y}(\m,v(\m))
         -\{\cdot,u(\m)\}^\varphi_\a
          \left[ {F(\m,v(\m))-\m^dc(\m)\over u(\m)}\right]_+ \cr
          & \quad - u(\m) \{\cdot,\left[ {F(\m,v(\m))-\m^dc(\m)\over u(\m)}
                         \right]_+ \}^\varphi_\a.\cr}\form$$
  Comparing (15) and (16), we obtain
    $$\eqalign{\m^a(\m^{d-1}X_{g_1}+\dots+X_{g_d})
       = & \left( \{\cdot,v(\m)\}^\varphi_\a{\partial F\over\partial
       y}(\m,v(\m))\right.\cr
         &\quad  \left. -\{\cdot,u(\m)\}^\varphi_\a
            \left[ {F(\m,v(\m))-\m^dc(\m)\over u(\m)}\right]_+ \right)
\moda{\a}
            u(\m),\cr}$$
  which proves (14).

  Finally, substituting
    $$G(\l) = \left[ {\l^dc(\l)\over u(\l)} \right]_+ u(\l)
              + F(\l,v(\l))\mod u(\l)$$
  in the expression
    $$\{G(\l),g_a\}^\varphi_\a = 0,$$
  we see that, for all $a=1,\dots,d$,
  $X_{g_a} c(\l)$ is completely determined by
  $X_{g_a}u(\l)$ and $X_{g_a}v(\l)$,
  which concludes the proof of the theorem.
\endproof

Our next step is to search for polynomials $\varphi'$ and
partitions $P'$ (associated to $\a'$), such that
the Hamiltonian vector fields
  $$X_{g_a} = \{\cdot,g_a\}^\varphi_\a, \quad a=1,\dots,d,$$
of the generalized master system $(M,\{\cdot,\cdot\}^\varphi_\a,g)$
coincide with the Hamiltonian vector fields of the system
$(M,\{\cdot,\cdot\}^{\varphi'}_{\a'},g')$, i.e.,
  $$\{\cdot,g_a\}^\varphi_\a = \{\cdot,g_a'\}^{\varphi'}_{\a'},$$
for all $a=1,\dots,d$.

We start by stating a simple lemma, which will turn out
to be very useful.
\lemma{Let $\a$ and $i$ be integers, $P(x)$ an arbitrary polynomial, and
  $u(x)=x^d+u_1x^{d-1}+\dots+u_d$ a generic polynomial.  Then
    $$\left[ {x^i P(x)\over u(x)}\right]_{\a+i}
      =x^i \left[ {P(x)\over u(x)}\right]_\a,\form$$
  and
    $$ (x^i P(x)) \moda{\a+i} u(x)= x^i (P(x)\moda{\a} u(x)).\form$$}
\proof  If
  $$P(x) = u(x) Q(x) + x^\a R(x),$$
  where $\deg R(x) < \deg u(x)$,
  then
    $$x^i P(x) = u(x) x^iQ(x) + x^{\a+i} R(x),$$
  and the result follows immediately.
\endproof

Let $\varphi$, $F(x,y,c)$ and $G(\l)$ be as above and
let $\a < n+1$ be an integer.
Further, we denote by
  $$P=\{\{f_0,\dots,f_n\},\{g_1,\dots,g_d\}\}$$
the partition of the set $I$ associated to $\a$ and we consider the
generalized master system $(M,\{\cdot,\cdot\}^\varphi_\a,g)$ associated
to this partition.  It is clear from Theorem 6 that
  $$\m^\a( \m^{d-1} X_{g_1}u(\l) + \dots + X_{g_d}u(\l) )
    = {\partial F\over\partial y}(\m,v(\m))
      \{u(\l),v(\m)\}^\varphi
      \moda{\a} u(\m).\form$$
Let $\varphi'$ be another polynomial, suppose $i$ is
an integer such that $\a+i\le n+1$ and denote by
  $$P'=\{\{f_0',\dots,f_n'\},\{g_1',\dots,g_d'\}\}$$
the partition of the set $I$, associated to $\a'=\a+i$.
Applying Theorem 6 for the generalized master system
$(M,\{\cdot,\cdot\}^{\varphi'}_{\a+i},g')$, we obtain
  $$\m^{\a+i}( \m^{d-1} X'_{g'_1}u(\l) + \dots + X'_{g'_d}u(\l) )
    = {\partial F\over \partial y}(\m,v(\m))
      \{u(\l),v(\m)\}^{\varphi'}
      \moda{\a+i} u(\m).\form$$
Our aim is to search for a polynomial $\varphi'$ such that the
Hamiltonian vector fields $X_{g_a}$ and $X'_{g'_a}$ coincide for all
$a=1,\dots,d$.  Comparing (19) and (20), we obtain the (necessary) condition
    $$ \m^i (\{u(\l),v(\m)\}^\varphi \moda{\a} u(\m))
      = \{u(\l),v(\m)\}^{\varphi'} \moda{\a+i} u(\m),\form$$
which, in view of Lemma 7, is equivalent to
    $$ \m^i \{u(\l),v(\m)\}^\varphi \moda{\a+i} u(\m)
      = \{u(\l),v(\m)\}^{\varphi'} \moda{\a+i} u(\m).\form$$
Using Proposition 1, we see that, for $a = 1,\dots,d$,
  $$\eqalignno{\m^i \{u_a,v(\m)\}^\varphi \moda{\a+i} u(\m)
             = & \m^i \{u(\m),v_a\}^\varphi \moda{\a+i} u(\m) \cr
             = & \m^i (\varphi(\m,v(\m))
                       \left[ {u(\m) \over \m^{d+1-a}} \right]_+
                       \mod u(\m) ) \moda{\a+i} u(\m) \cr
             = & \m^i \varphi(\m,v(\m))
                      \left[ {u(\m) \over \m^{d+1-a}} \right]_+
                      \moda{\a+i} u(\m)\cr
             = & (\m^i \varphi(\m,v(\m))
                  \left[ {u(\m) \over \m^{d+1-a}} \right]_+ \mod u(\m) )
                  \moda{\a+i} u(\m) \cr
             = & \{u(\m),v_a\}^{x^i \varphi} \moda{\a+i} u(\m)\cr
             = & \{u_a,v(\m)\}^{x^i \varphi} \moda{\a+i} u(\m),
             &\formula\cr}$$
and hence,
  $$\m^i\{u(\l),v(\m)\}^\varphi \moda{\a+i} u(\m)
    = \{u(\l),v(\m)\}^{x^i\varphi} \moda{\a+i} u(\m).\form$$
Comparison of the expressions (22) and (24) suggests taking
$\varphi'=x^i\varphi$, and we obtain the following
\theorem{Let $\a$ and $i$ be integers such that $\a+i\le n+1$, and let
    $$P=\{\{f_0,\dots,f_n\},\{g_1,\dots,g_d\}\}
      \quad\hbox{\sl and }\quad
      P'=\{\{f_0',\dots,f_n'\},\{g_1',\dots,g_d'\}\}$$
  be the partitions of the set $I$, associated to $\a$ and $\a'=\a+i$.  Then
    $$\m^i(\{u(\l),v(\m)\}^\varphi \moda{\a} u(\m))
    = \{u(\l),v(\m)\}^{x^i\varphi} \moda{\a+i} u(\m),\form$$
  and the Hamiltonian vector fields of the generalized master systems
  $(M,\{\cdot,\cdot\}^\varphi_\a,g)$
  and $(M,\{\cdot,\cdot\}^{x^i\varphi}_{\a+i},g')$ coincide, i.e.,
    $$\{\cdot,g_a\}^\varphi_\a = \{\cdot,g_a'\}^{x^i\varphi}_{\a+i}$$
  for all $a=1,\dots,d$.}
\proof
  The first statement follows immediately from (24) and
  Lemma 7.  Using Theorem 6 and (25), we see that
    $$\eqalign{\{u(\l),g_a'\}^{x^i\varphi}_{\a+i}
                = &\{u(\l),g_a\}^\varphi_\a,\cr
               \{v(\l),g_a'\}^{x^i\varphi}_{\a+i}
                = &\{v(\l),g_a\}^\varphi_\a,\cr}\form$$
  for all $a=1,\dots,d$.  As the Hamiltonian vector fields $X_{g_a}$ and
  $X'_{g_a'}$ are
  completely determined by (14), this proves the second part of the theorem.
\endproof

As a consequence of this theorem, we can write the Hamiltonian vector fields
of the generalized master system
$(M,\{\cdot,\cdot\}^\varphi_\a,g)$ in Hamiltonian form
with respect to a set of Poisson brackets
$\{\cdot,\cdot\}^{x^i\varphi}_{\a+i}$, indexed by the integers $i$
such that $\a+i\le n+1$.
We will now show that all these Poisson brackets are compatible.
To this purpose, it suffices to show that the Poisson brackets are pairwise
compatible, i.e., that for any polynomial $\varphi$ and
any integers $\a$ and $i$ such that $\a+i\le n+1$, the Poisson brackets
$\{\cdot,\cdot\}^\varphi_\a$ and $\{\cdot,\cdot\}^{x^i\varphi}_{\a+i}$
are compatible.
To prove this assertion, we will use a very interesting trick
which was, to our knowledge, first used by P. Vanhaecke in [Va1],
to show that the Poisson brackets
of a two-dimensional completely bi-Hamiltonian integrable system
(we refer to [Va1] for the exact definition of such systems) are compatible.

\theorem{Let $\varphi$ be an arbitrary polynomial, and suppose that
  $\a$ and $i$ are integers such that $\a+i\le n+1$.  Further, we denote by
    $$P=\{\{f_0,\dots,f_n\},\{g_1,\dots,g_d\}\}
      \quad\hbox{\sl and }\quad
      P'=\{\{f_0',\dots,f_n'\},\{g_1',\dots,g_d'\}\}$$
  the partitions of the set $I$ associated to $\a$ and $\a'=\a+i$.
  Then the Poisson brackets $\{\cdot,\cdot\}^\varphi_\a$ and
  $\{\cdot,\cdot\}^{x^i\varphi}_{\a+i}$ of the generalized master
  systems $(M,\{\cdot,\cdot\}^\varphi_\a,g)$
  and $(M,\{\cdot,\cdot\}^{x^i\varphi}_{\a+i},g')$ are compatible.}
\proof
  To simplify the notation, we denote
  $\{\cdot,\cdot\}_1=\{\cdot,\cdot\}^\varphi_\a$ and
  $\{\cdot,\cdot\}_2=\{\cdot,\cdot\}^{x^i\varphi}_{\a+i}$.  Further,
  we write the polynomial $G(\l)$ as
    $$G(\l) = H_0\l^{d+n}+\dots+H_{d+n},$$
  and we denote by
    $$\{\cdot,\cdot\}=\{\cdot,\cdot\}_1+\{\cdot,\cdot\}_2$$
  the sum of the two brackets.

  To prove the compatibility of $\{\cdot,\cdot\}_1$ and
  $\{\cdot,\cdot\}_2$, it suffices to prove the Jacobi identity
    $$\{\{f,g\},h\}+\{\{g,h\},f\}+\{\{h,f\},g\} = 0\form$$
  for all functions $f,g,h \in
  \{u_1,\dots,u_d,v_1,\dots,v_d,H_0,\dots,H_{d+n}\}$.
  Using the fact that $\{\cdot,\cdot\}_1$ and $\{\cdot,\cdot\}_2$ are Poisson
  brackets and hence satisfy the Jacobi identity, (27) reduces to
  the ``mixed Jacobi identity''
    $$\eqalign{
      {}&\{\{f,g\}_2,h\}_1+\{\{g,h\}_2,f\}_1+\{\{h,f\}_2,g\}_1\cr
      {}&\quad+\{\{f,g\}_1,h\}_2+\{\{g,h\}_1,f\}_2+\{\{h,f\}_1,g\}_2 = 0.\cr}
      \form$$
  From Proposition 1 we know that $\{\cdot,\cdot\}^\varphi$
  and $\{\cdot,\cdot\}^{x^i\varphi}$ are compatible.  Hence, (28) holds if
  $f,g,h \in \{u_1,\dots,u_d,v_1,\dots,v_d\}$, and we only have to prove
  (28) in the case where at least one function (say, $h$,) belongs to
  $\{H_0,\dots,H_{d+n}\}$.  As Theorem 8 implies that, for all $i =
  0,\dots,d+n$, we have
    $$\eqalignno{\{\cdot,H_i\}_2 & = \{\cdot,H_j\}_1&\formula\cr
                 \{\cdot,H_i\}_1 & = \{\cdot,H_k\}_2&\formula\cr}$$
  for some $j$ and $k$, we see that
    $$\eqalign{
      {}&\{\{f,g\}_2,H_i\}_1+\{\{g,H_i\}_2,f\}_1+\{\{H_i,f\}_2,g\}_1\cr
      {}&\quad+\{\{f,g\}_1,H_i\}_2+\{\{g,H_i\}_1,f\}_2+\{\{H_i,f\}_1,g\}_2\cr
      = &\{\{f,g\}_2,H_k\}_2+\{\{g,H_j\}_1,f\}_1+\{\{H_j,f\}_1,g\}_1\cr
      {}&\quad+\{\{f,g\}_1,H_j\}_1+\{\{g,H_k\}_2,f\}_2+\{\{H_k,f\}_2,g\}_2\cr
      = & 0,\cr}$$
  as $\{\cdot,\cdot\}_1$ and $\{\cdot,\cdot\}_2$ satisfy the Jacobi
  identity.
\endproof

Combining the results stated in this section, we obtain the following
\theorem{Let $F(x,y,c)=F(x,y)-x^d c(x)$
  be a deformation family of polynomials, $\varphi$ an arbitrary
  non-zero polynomial, and
    $$G(\l) = \l^d c(\l) + (F(\l,v(\l))-\l^d c(\l))\mod u(\l)$$
  the polynomial whose coefficients are given by the set
    $$I=\{c_0,\dots,c_n,H_1,\dots,H_d\}$$
  of invariants of the system
  $(\R^{2d+n+1},\{\cdot,\cdot\}^\varphi_c,h)$.  Further, suppose that
  $\a< n+1$ is an integer and let
    $$P=\{\{f_0,\dots,f_n\},\{g_1,\dots,g_d\}\}$$
  be the partition of the set $I$ associated to $\a$,
  i.e., the partition given by
    $$G(\l) = f_0 \l^{d+n} + \dots + f_{n-\a}\l^{d+\a}
              + \l^\a(g_1\l^{d-1}+\dots +g_d)
              + f_{n-\a+1} \l^{\a-1}+\dots +f_n.$$
  The generalized master system $(M,\{\cdot,\cdot\}^\varphi_\a,g)$
  admits a multi-Hamiltonian structure, which is constructed as follows.
  For every integer $\a'=\a+i\le n+1$, the partition
    $$P'=\{\{f_0',\dots,f_n'\},\{g_1',\dots,g_d'\}\}$$
  of the set $I$, associated to $\a'$, determines a
  Poisson bracket $\{\cdot,\cdot\}^{x^i\varphi}_{\a+i}$, and the
  Hamiltonian vector fields $\{\cdot,g_a\}^\varphi_\a$, $a=1,\dots,d$,
  of the system can be written in Hamiltonian form
      $$\{\cdot,g_a\}^\varphi_\a=\{\cdot,g_a'\}^{x^i\varphi}_{\a+i},$$
  with respect to the (compatible) Poisson bracket
  $\{\cdot,\cdot\}^{x^i\varphi}_{\a+i}$.}

\remark  Strictly speaking, it follows from Theorem 8 that well-chosen
families of generalized master systems
have the same Hamiltonian vector fields (and will therefore not be
considered as {\sl different} systems in the strict sense).
Combining this result with Theorem 9,
we see that the class of generalized master systems therefore consists of
a (smaller) class of completely integrable systems, together with the
systems determined by their multi-Hamiltonian structure, and from this
stricter point of view, the construction given in our paper actually
yields a multi-Hamiltonian formulation for a smaller class of completely
integrable systems (instead of, as was stated in Theorem 10, for all
generalized master systems).

\remark  In Corollary 5, we have constructed a class of
generalized master systems, starting from a (special) partition
  $$P=\{\{f_0,\dots,f_n\},\{g_1,\dots,g_d\}\}$$
of the set $I$, associated to an integer $\a\le n+1$, i.e., determined by
    $$G(\l) = f_0 \l^{d+n} + \dots + f_{n-\a}\l^{d+\a}
              + \l^\a(g_1\l^{d-1}+\dots +g_d)
              + f_{n-\a+1} \l^{\a-1}+\dots +f_n.$$
Choosing an {\sl arbitrary} partition
  $$P=\{\{f_0,\dots,f_n\},\{g_1,\dots,g_d\}\}$$
of the set $I$, the mapping $\phi$ (associated to $f$)
is again bijective on a dense set $M$ in $\R^{2d+k}$,
and applying Theorem 4 in this case, we again obtain
a completely integrable system on $M$, which will also be called a {\it
generalized master system}.  In the rest of this section,
we will show how to adapt our construction of a multi-Hamiltonian
formulation to the case of arbitrary partitions.

We start by introducing some terminology, similar to that in Section 2.
Let $P(x)$ be an arbitrary polynomial and assume that $u(x) =
x^d+u_1x^{d-1}+\dots+u_d$ is a {\sl generic} polynomial.  Further, let
$\a=(a_1,\dots,a_d)$ be a fixed $d$-tuple of different integers.
The polynomial $P(x)$ can be written, in a unique way, as
  $$P(x) = Q(x) u(x) + \sum_{i=1}^d R_i x^{a_i},$$
and as in Section 2 we use the notation
  $$Q(x) = \left[ {P(x) \over u(x)} \right]_\a,\quad
    R(x) =\sum_{i=1}^d R_i x^{a_i} = P(x) \moda{\a} u(x).$$

An arbitrary partition
  $$ P=\{\{f_0,\dots,f_n\},\{g_1,\dots,g_d\}\}$$
of the set
  $$I=\{c_0,\dots,c_n,H_1,\dots,H_d\}$$
corresponds to a $d$-tuple $\a=(a_1,\dots,a_d)$,
$g_j$ being the
coefficient of $x^{a_j}$ in $G(\l)$.
(We remark that this $d$-tuple satisfies the additional condition that
$\a\le d+n$, i.e., $a_j\le d+n$ for all $j=1,\dots,d$.)
Using this correspondence between
partitions and $d$-tuples, together with the notation introduced before,
the expressions for the Hamiltonian vector fields
given in Theorem 6 remain valid in the case of a general partition.

To construct the multi-Hamiltonian formulation in this more general
case, we consider two partitions
  $$ P=\{\{f_0,\dots,f_n\},\{g_1,\dots,g_d\}\}
     \quad \hbox{and} \quad
     P'=\{\{f_0',\dots,f_n'\},\{g_1',\dots,g_d'\}\}$$
of the set $I$, corresponding to two
$d$-tuples $\a=(a_1,\dots,a_d)$ and $\a+i=(a_1+i,\dots,a_d+i)$.
One can then show (by the same argument as before)
that the Hamiltonian vector fields
of the generalized master systems
$(M,\{\cdot,\cdot\}^\varphi_f,g)$ and
$(M,\{\cdot,\cdot\}^{x^i\varphi}_{f'},g')$ coincide,
and that the Poisson brackets $\{\cdot,\cdot\}^\varphi_f$
and $\{\cdot,\cdot\}^{x^i\varphi}_{f'}$ are again compatible,
which concludes our construction.


\section{Multi-Hamiltonian formulation for the odd master system}


This last section is devoted to the study of a simple but very
illustrative example of a (two-dimensional) generalized master system.
More precisely, we consider the completely integrable system
$(\R^7,\{\cdot,\cdot\}^1_0,g)$, associated to the polynomial
  $$F(x,y) = y^2+f(x),\quad \deg f(x)=5.$$
We  compute the Hamiltonian vector fields of this system,
and we construct a multi-Hamiltonian formulation for this
system by using the theory developed in the previous section.

We start our investigation by making some notational conventions.
We put $d=2$, $n=2$, and we denote
  $$\eqalign{ u(\l) & = \cr v(\l) & = \cr c(\l) & = \cr}
    \eqalign{\l^2+&u_1\l+u_2,\cr
                  &v_1\l+v_2,\cr
          c_0\l^2+&c_1\l+c_2.\cr}$$
Further, we fix a polynomial $F(x,y)$, given by
  $$F(x,y) = f(x) + y^2, \quad f(x) = x^5+Ax^4+Bx^3+Cx^2+Dx+E.$$
Finally, for reasons that will become clear at the end of this section,
we replace the
variables $c=(c_0,c_1,c_2)$ by a new set of variables $(w_1,w_2,w_3)$,
given by
  $$\eqalign{ w(\l) = & \l^3+w_1\l^2+w_2\l+w_3\cr
                    = & \left[ {F(\l,v(\l)) - \l^2 c(\l)\over u(\l)}
                    \right]_+.\cr}$$
In these new variables, the polynomial
  $$\eqalign{G(\l) = & \l^2 c(\l) + (F(\l,v(\l))-\l^2 c(\l))\mod u(\l)\cr
                   = & F(\l,v(\l)) - u(\l) w(\l)\cr}$$
takes the form
  $$G(\l) = H_1\l^4+H_2\l^3+H_3\l^2+H_4\l+H_5,$$
where
  $$\eqalign{H_1&=A-u_1-w_1,\cr
             H_2&=B-u_2-u_1w_1-w_2,\cr
             H_3&=C+v_1^2-u_2w_1-u_1w_2-w_3,\cr
             H_4&=D+2v_1v_2-u_2w_2-u_1w_3,\cr
             H_5&=E+v_2^2-u_2w_3.\cr}$$

We will now compute the Poisson bracket $\{\cdot,\cdot\}^1_0$ associated
to the partition
  $$P_0=\{f=\{H_1,H_2,H_3\},g=\{H_4,H_5\}\}.$$
Putting $\varphi=1$
in Proposition 1, we see that the Poisson matrix of the bracket
$\{\cdot,\cdot\}^1$ (on $\R^4$) is of the form
  $$A=\pmatrix{0 & 0 & 0 & 1   \cr
               0 & 0 & 1 & u_1 \cr
               0 & -1& 0 & 0   \cr
              -1 &-u_1 & 0 & 0 \cr}.$$
To compute the remaining components of the Poisson bracket
$\{\cdot,\cdot\}^1_0$, we use the following simple method:
requiring $H_1$, $H_2$ and $H_3$ to be Casimir
functions of the Poisson bracket, i.e., $\{\cdot,H_i\}^1_0=0$ for
$i=1,2,3$, we obtain a system of equations in the components
of the Poisson bracket, which is then solved for the unknown
components.  Using this method, we find that
the Poisson bracket $\{\cdot,\cdot\}^1_0$ is given (with respect to the
coordinates $(u,v,w)$) by
  $$\pmatrix{0 & 0 & 0 & 1   & 0 & 0 & 0   \cr
             0 & 0 & 1 & u_1 & 0 & 0 & 2v_1\cr
             0 & -1& 0 & 0   & 0 & 1 & w_1-u_1\cr
            -1 &-u_1 & 0 & 0 & 1 & w_1 & w_2-u_2\cr
             0 & 0 & 0 & -1 & 0 & 0 & 0\cr
             0 & 0 & -1 & -w_1 & 0 & 0 & -2v_1\cr
             0 & -2v_1 & u_1-w_1 & u_2-w_2 & 0 & 2v_1 & 0\cr}.
             \form$$

The Hamiltonian vector fields $X_{H_4}$ and $X_{H_5}$
of the system $(\R^7,\{\cdot,\cdot\}^1_0,g)$
take the following form (this can be computed directly from (31), or by
using Theorem 6):
  $$X_{H_4}\left\{
    \eqalign{\dot{u}_1&=2v_1,\cr
             \dot{u}_2&=2v_2,\cr
             \dot{v}_1&=w_2-u_2-u_1w_1+u_1^2,\cr
             \dot{v}_2&=w_3-w_2u_1+u_1u_2,\cr
             \dot{w}_1&=-2v_1,\cr
             \dot{w}_2&=-2v_2-2v_1w_1+2v_1u_1,\cr
             \dot{w}_3&=2u_1v_2-2v_2w_1,\cr}\right.
    \qquad
    X_{H_5}\left\{
    \eqalign{u_1'&=2v_2,\cr
             u_2'&=2v_2u_1-2v_1u_2,\cr
             v_1'&=w_3-u_2w_1+u_1u_2,\cr
             v_2'&=u_1w_3-w_2u_2+u_2^2,\cr
             w_1'&=-2v_2,\cr
             w_2'&=-2v_2w_1+2v_1u_2,\cr
             w_3'&=2v_1w_3+2u_2v_2-2v_2w_2.\cr}\right.\form$$

Using the theory developed in Section 4, the Hamiltonian vector fields
(32) of this system can be written in Hamiltonian form with respect to
three additional (compatible) Poisson brackets
$\{\cdot,\cdot\}^{x^i}_i$, $i=1,2,3$.
We will now compute these brackets explicitly.

First, putting $\varphi=x$ in Proposition 1, the Poisson matrix
of the bracket $\{\cdot,\cdot\}^x$ takes the form
  $$\pmatrix{0 & 0 & 1 & 0    \cr
             0 & 0 & 0 & -u_2 \cr
             -1 & 0 & 0 & 0   \cr
             0 & u_2 & 0 & 0  \cr}.$$
Using the fact that $H_1$, $H_2$ and $H_5$ are Casimir functions of the
bracket $\{\cdot,\cdot\}^x_1$, its Poisson matrix takes the form
  $$\pmatrix{0 & 0 & 1 & 0 & 0 & 0 & 0\cr
             0 & 0 & 0 & -u_2 & 0 & 0 & -2v_2\cr
             -1 & 0 & 0 & 0 & 1 & w_1-u_1 & 0\cr
             0 & u_2 & 0 & 0 & 0 & -u_2 & -w_3\cr
             0 & 0 & -1 & 0 & 0 & 0 & 0\cr
             0 & 0 & u_1-w_1 & u_2 & 0 & 0 & 2v_2\cr
             0 & 2v_2 & 0 & w_3 & 0 & -2v_2 & 0\cr},$$
and it is clear (from Theorem 8 or by direct computation)
that $X_{H_4} = \{\cdot,H_3\}^x_1$ and $X_{H_5}=\{\cdot,H_4\}^x_1$,
showing the Hamiltonian form of the Hamiltonian
vector fields (32) with respect to this second bracket.
A similar computation yields the following Poisson matrix for the
Poisson bracket $\{\cdot,\cdot\}^{x^2}_2$:
  $$\pmatrix{0 & 0 & -u_1 & -u_2 & 0 & -2v_1 & -2v_2\cr
             0 & 0 & -u_2 & 0 & 0 & -2v_2 & 0\cr
             u_1 & u_2 & 0 & 0 & -u_1 & -w_2 & -w_3\cr
             u_2 & 0 & 0 & 0 & -u_2 & -w_3 & 0\cr
             0 & 0 & u_1 & u_2 & 0 & 2v_1 & 2v_2\cr
             2v_1 & 2v_2 & w_2 & w_3 & -2v_1 & 0 & 0\cr
             2v_2 & 0 & w_3 & 0 & -2v_2 & 0 & 0\cr}.$$
(Again, it is easily checked that $X_{H_4}$ and $X_{H_5}$ can be written
in Hamiltonian form with respect to this new bracket.)
Finally, putting $\varphi=x^3$ in Proposition 1, we see that
$\{\cdot,\cdot\}^{x^3}$ is given by the matrix
  $$A=\pmatrix{0 & 0 & u_1^2-u_2 & u_1u_2   \cr
               0 & 0 & u_1u_2 & u_2^2       \cr
               -u_1^2+u_2 & -u_1u_2 & 0 & 0 \cr
               -u_1u_2 & -u_2^2 & 0 & 0     \cr}.$$
Requiring $H_3$, $H_4$ and $H_5$ to be the Casimir functions of the bracket
$\{\cdot,\cdot\}^{x^3}_3$, we see that its Poisson matrix is of the form
  $$\pmatrix{A&B\cr -{}^tB & C\cr},$$
where
  $$ B=\pmatrix{-2v_1 & 2u_1v_1-2v_2 & 2v_2u_1\cr
                -2v_2 & 2u_2v_1 & 2v_2u_2\cr
                -w_2+w_1u_1 &-w_3+u_1w_2 & u_1w_3\cr
                -w_3+w_1u_2 & u_2w_2 & u_2w_3\cr},$$
and
  $$ C=\pmatrix{0 & -2v_1w_1 & -2v_2w_1\cr
               2v_1w_1 & 0 & 2v_1w_3-2v_2w_2\cr
               2v_2w_1 & 2v_2w_2-2v_1w_3 & 0\cr},$$
and we can again write the vector fields (32) in Hamiltonian form with
respect to this bracket.

\remark  In [M], D. Mumford constructs
a natural vector field on the Jacobian of
any hyperelliptic curve of genus $g$, given by an equation
$y^2=f(x)$, where $\deg f(x) = 2g+1$.  This vector field turns out to
determine a $g$-dimensional completely integrable Hamiltonian system.
The construction of these integrable systems was later adapted
to the case where $\deg f(x) = 2g+2$ by P. Vanhaecke [Va1],
who also made a detailed study of these systems,
which he called the {\it odd} (resp. {\it even}) {\it master systems},
in the two-dimensional case (i.e., $g=2$).

The existence of a multi-Hamiltonian formulation for the two-dimensional
master systems was already shown by P. Vanhaecke [Va1].
However, his construction was of a heuristic nature, and the
computation of the compatible Poisson brackets required a
trial-and-error method.

Comparing the differential equations (32) to the expressions given in
[Va1],  we see that the two-dimensional odd master system and the system
$(\R^7,\{\cdot,\cdot\}^1_0,g)$ given by (32) are the same (up to a
constant factor), which provides us with a (more systematic)
construction of the multi-Hamiltonian
formulation for the (two-dimensional) odd master
system given in [Va1].  A similar computation, starting from the polynomial
  $$F(x,y)=y^2+x^6+Ax^5+Bx^4+Cx^3+Dx^2+Ex+F,$$
yields a multi-Hamiltonian formulation for the (two-dimensional) even master
system.

\references

\item{[AR1]}
      M. Antonowicz and S. Rauch-Wojciechowski,
      Bi-Hamiltonian formulation of the H\'enon-Heiles system and its
      multi-dimensional extensions,
      {\it Phys.\ Lett.\ A} 163 (1992), 167--172.
\item{[AR2]}
      M. Antonowicz and S. Rauch-Wojciechowski,
      How to construct finite-dimensional bi-Hamiltonian systems from
      soliton equations: Jacobi integrable potentials,
      {\it J.\ Math.\ Phys.} 33 (1992), 2115--2125.
\item{[B]}
      R. Brouzet, Syst\`emes bihamiltoniens et compl\`ete
      int\'egrabilit\'e en dimension 4,
      {\it C.\ R.\ Acad.\ Sci.\ Paris S\'er.\ I} 311 (1990), 895--898.
\item{[BMT]}
      R. Brouzet, P. Molino and F. Turiel,
      G\'eom\'etrie des syst\`emes bihamiltoniens,
      {\it Indag.\ Mathem.\ (N.S.)} 4 (1993), 269--296.
\item{[BR]}
      M. Bruschi and O. Ragnisco,
      On a new integrable Hamiltonian system with nearest-neighbour
      interaction, {\it Inverse Problems} 5 (1989), 983--998.
\item{[CRG]}
      R. Caboz, V. Ravoson and L. Gavrilov,
      Bi-Hamiltonian structure of an integrable H\'enon-Heiles system,
      {\it J.\ Phys.\ A} 24 (1991), L523--L525.
\item{[D]}
      P. Damianou, Master symmetries and $R$-matrices for the Toda
      lattice, {\it Lett.\ Math.\ Phys.} 20 (1990), 101--112.
\item{[MM]}
      G. Magnano and F. Magri, Poisson-Nijenhuis structures and Sato
      hierarchy, {\it Rev.\ Math.\ Phys.} 3 (1991), 403-466.
\item{[Ma]}
      F. Magri, A simple model of the integrable Hamiltonian equation,
      {\it J.\ Math.\ Phys.} 19 (1978), 1156--1162.
\item{[M]}
      D. Mumford, {\it Tata lectures on theta II},
      Birkh\"auser, Boston-Basel-Stuttgart, 1984.
\item{[OR]}
      W. Oevel and O. Ragnisco,
      $R$-matrices and higher Poisson brackets for integrable systems,
      {\it Phys.\ A} 161 (1989), 181--220.
\item{[R]}
      V. Ravoson,
      $(r,s)$-structure bi-Hamiltonienne, s\'eparabilit\'e, paires de Lax
      et int\'egrabilit\'e,
      Th\`ese de doctorat, Universit\'e de Pau et des Pays de l'Adour,
      1992.
\item{[Va1]}
      P. Vanhaecke, Linearising two-dimensional integrable systems and the
      construction of action-angle variables,
      {\it Math.\ Z.} 211 (1992), 265-313.
\item{[Va2]}
      P. Vanhaecke, Integrable systems and symmetric products of curves,
      {\it Pub.\ IRMA Lille} 33 (1993).

\end